\documentclass[prd,superscriptaddress,onecolumn,showpacs,11pt,msmath,preprintnumbers,showkeys]{revtex4}
\usepackage{wrapfig,rotating}
\usepackage{graphicx,epsfig,color}
\usepackage{appendix}

\makeatletter

\newenvironment{figurehere}
{\def\@captype{figure}}
{}
\makeatother

\usepackage{graphicx}
\usepackage{dcolumn}
\usepackage{bm}

\def\beq{\begin{equation}}
\def\eeq{\end{equation}}
\def\beeq{\begin{eqnarray}}
\def\eeeq{\end{eqnarray}}

\def\2GPD{$_2\mbox{GPD}$}

\def\12{$1\otimes 2$}
\def\22{$2 \otimes 2$}

\def\Qsep{Q_{\mbox{\rm\scriptsize sep}}}
\def\Qsep2{Q^2_{\mbox{\rm\scriptsize sep}}}

\begin{document}
 \title{Dynamic scaling and quenching for heavy quark in the linear expanding medium.}
 \author{B.\ Blok,
C.\ Wu \\[2mm] \normalsize  Department of Physics, Technion -- Israel Institute of Technology,
 Haifa, Israel}
 \begin{abstract}
 We study  the energy loss of the heavy
quark and the heavy quark jet propagating through linear expanding quark-gluon plasma (QGP).
We show that the concept of scaling is valid for heavy quarks, with the quenching parameter of the
equivalent static media being independent from the mass of the quark.
 \end{abstract}

   \maketitle
 \thispagestyle{empty}

 \vfill

\section{Introduction.}
\par 
The aim of this paper is to study the energy loss of the heavy quark and the heavy quark jet in expanding dense media in the Baier-Dokshitzer-Mueller-Peigne-Schiff -Zakharov  (BDMPSZ)  \cite{BDMPS1,BDMPS2,BDMPS3,BDMPS4,Z1,Z2} approach
(see  Ref. \cite{MT1} for a recent review), based on taking  into account the  
Landau-Pomeranchuk -Migdal  (LPM) effect for  the parton propagating through the quark gluon plasma (QGP).  
\par The energy loss in the  expanding media in comparison to the static one has attracted a lot of attention for massless  partons, starting from \cite{BDMPS5,Arnold,SW1,SW2}, see \cite{T4,T3,T1,T2} for recent studies.  The basic observation was the so called 
scaling phenomena \cite{SW1,SW2}: the energy loss of the massless parton in the expanding  media  is given by the same function $f(\omega/\omega_c)$, where $\omega_c$ is the characteristic frequency 
(different for expanding and static media, see below), as in the static media with the effective static quenching coefficient 
$\hat q_{\rm stat}$. Similar ideas were discussed for the energy loss of the gluon and massless quark  jets \cite{T4}.
\par On the other hand a lot of attention was directed to the problem of the energy loss of the heavy quark propagating  through the dense static media
\cite{DK,ASW,Z3,Z4,B1,B2,B3}. 
Nevertheless, since the idea of scaling was proposed in \cite{SW1,SW2}, the relevant calculations were carried out only for  the massless case.
 Then  it is interesting what happens with scaling for massive case, in particular whether the  quenching coefficient of the equivalent static media $\hat q_{\rm stat}$ depends on the heavy quark mass or the dead cone angle.
In  this note  we show that  both Salgado--Wiedemann scaling \cite{SW1,SW2}  and Caucal--Iancu--Soyez  scaling \cite{T3,T4} work independent of mass for the energy loss in the 
the linearly expanding media. We also calculate quenching weights for  the  expanding media and see that they 
depend only weakly on heavy quark mass for media induced (MIE) emissions, and are very close numerically to the corresponding quenching weights of  the equivalent static media. 
\par We shall consider here the physically most interesting case of the Bjorken linear expansion.
\par In this paper we call BDMPSZ  the mean field approach developed in \cite{BDMPS1,BDMPS2,BDMPS3,BDMPS4,Z1,Z2}.
In these papers the kinematical constraint $k_t<\omega$ was neglected ( where $k_t$ is the transverse momenta, and $\omega$ is the energy
of the radiated parton), the  integration in $k_t$  was carried till the infinity limit.  It was shown in \cite{ASW} how to take into account the constraint $k_t<\omega$, this constraint is especially important for heavy quark radiation.  We shall call
the mean field approach with the kinematical constraint $k_t<\omega$ included the Armesto-Salgado-Wiedemann (ASW approach.
\par The paper is organized in the following way. In section 2 we briefly review the basic formalism for  calculation of energy 
loss of the heavy quark in the expanding media. In section 3 we present the formulae for calculating bulk and boundary parts of the heavy quark energy loss. In section 4 we study the Salgado--Wiedemann scaling and compare the quenching weights in
expanding and equivalent static media. In section 5 we study the Caucal--Iancu--Soyes scaling for the energy loss of the heavy quark jets. Our results are summarised in Section 6.
\section{Basic formalism for the calculation of the heavy quark energy loss in the expanding media.} 
\par Let us briefly recall some  details about ASW and BDMPSZ approaches in the expanding media needed for our calculation
\cite{BDMPS5,SW1}.
\par In the ASW approach we start from the expression for heavy quark radiation  in terms of the heavy quark propagator:
\begin{eqnarray}
\omega\frac{dI}{d\omega d^2k_t}&=&\frac{\alpha_sC_F}{(2\pi)^2
\omega^2}2Re \int _{t_0}^\infty dt\int_{t}^\infty dt _1\int d^2u\,e^{-i\vec{k}_{t}\vec{u}}\\[10pt]
&\times&\exp{[-\frac{1}{2}\int^\infty_{t_1}ds n(s)\sigma (\vec u)]}
\frac{\partial}{\partial\vec u}\frac{\partial}{\partial\vec y }K(\vec u, t_1;\vec y, 0)_{\vec y=\vec 0},\nonumber,
\label{one1}
\end{eqnarray}
where the heavy quark propagator K is given for the expanding media by 
\beq
K(\vec r_1,t_1;\vec r_2 t)=
\frac{\omega}{2\pi D(t_1,t)}\exp{\frac{i\omega}{2 D(t_1,t)}(c_1r_1^2+c_2r_2^2-2\vec r_1\vec r_2)}
e^{-i(t_1-t)\theta_0^2\omega/2}.
\label{2}
\eeq
Here 
\begin{eqnarray}
D(t_1,t)&=&\frac{2}{2-h}\sqrt{tt_1}
\left[I_{\nu}(U(t_{1}))K_{\nu}(U(t))-I_{\nu}(U(t))K_{\nu}(U(t_{1}))\right],\\[10pt]
c_1(t_1,t)&=&\frac{1}{2-h}\sqrt{tt_1}(1+i)\sqrt{\frac{\hat q_{0}}{\omega}}\left (\frac{t_{0}}{t_1}\right)^{h/2} \left[I_{\nu-1}(U(t_1))K_{\nu}(U(t))+I_{\nu}(U(t))K_{\nu-1}(U(t_1)\right],\\[10pt]
c_2(t_1,t)&=&c_1(t,t_1).
\label{3}
\end{eqnarray}
Here $\omega$ and $\vec k_t$ are the energy and the transverse momentum of the radiated parton, I and K are the modified Bessel functions.
The function U(t) is given by
\beq
U(t)=\frac{1}{2-h}\sqrt{\frac{\hat q_0}{\omega}}(1+i)t(\frac{t_0}{t})^{h/2}.
\eeq
The dependence of the density and the quenching coefficient on time  are given by
\beq
\hat q(t)=\hat q_0(\frac{t_{0}}{t})^{h},\,n(t)=n_0(\frac{t_{0}}{t})^{h},
\eeq
and 
\beq
\nu=\frac{1}{2-h}.
\eeq
It is easy to check that in the limit $h \rightarrow 0$ we get the static media, while for $h \rightarrow 1$ we have the Bjorken linear expansion. The parameter $t_0$ corresponds to the start of the plasma expansion ,i.e.
$t_0=0.1-0.2 $\,fm $=0.5-1 $\,GeV$^{-1}$,  while for $\hat q_0$ we take $3-6$ GeV$^3$ that is obtained from phenomenological estimates for Bjorken expansion \cite{T4}. The parameter  $\theta_0=m/E$  is the dead cone angle.
We use the Gyulassy--Wang model for quark-gluon plasma (QGP). We shall work in the harmonic approximation,
and consider the "brick" model of QGP, so the effective transverse cross section is 
\beq
\sigma (\vec u)=\frac{1}{2}\hat q(t)u^2.
\eeq
 \section{Energy loss  of the heavy quark in the expanding media and the quenching weights.}
\par If we explicitly integrate  Eq.1 
 over $k_t$  imposing phase space constraints, we get the ASW expression for the energy loss of the heavy quark.
Conventionally  we use  the kinematic constraint $k_t\le \omega$.
\par The energy loss and the gluon distribution in the media can be split into the boundary and the bulk contributions.
The first contribution arises  when the gluon is radiated from the media and absorbed in the vacuum, while the second contribution 
is when the gluon is both radiated and absorbed by the media.
Although such splitting  can be avoided for massless case using the simple approach of \cite{Arnold},  for massive quark we are forced to take into account each of these two contributions separately. 
\par We can also neglect phase constraints on transverse momenta and integrate over $k_t$ till infinity. In this case we get 
the original BDMPSZ approach. In the latter case we can regularise the obtained expressions  by subtracting explicitly the vacuum contribution.
For ASW case integrating Eq. \ref{one1} over $k_t$ up to $k_t=\omega_1$ we obtain explicitly:
 \begin{eqnarray}
  \omega\frac{dI^{\rm bulk}}{d\omega}&=&-\frac{\alpha_sC_F}{\pi}{\rm Re}\int _{t_0}^{L+t_0} dt\int_{t}^{L+t_0} dt _1
  \frac{2}{D^{2}(t_{1},t)^{2}}e^{-i\theta_0^{2}\omega(t_{1}-t)/2}\nonumber\\[10pt]
  &\times&\left(1-e^{-\omega_{1}^{2}/4F}+\frac{i\omega c_{1}(t_{1},t)\omega_{1}^{2}e^{-\omega_{1}^{2}/4F}}{8D(t_{1},t)F^{2}}\right),\nonumber\\[10pt]
  \label{10}
  \end{eqnarray}
  where  for linear expansion (h=1)
  \begin{equation}
  F(t_1,t)=\frac{1}{4}\hat q_0\log((L+t_0)/t_1)-\frac{i\omega c(t_1,t)}{2D(t_1,t)},
  \end{equation}
  and in general case 
  \begin{equation}
  F(t1_1,t)=-\frac{1}{4(1-h)}\hat q_0(1/t_0^{h-1}-1/(L+t_0)^{h-1})-\frac{i\omega c(t_1,t)}{2D(t_1,t)}.
  \end{equation}
   \begin{eqnarray}
  \omega\frac{dI^{\rm boundary}}{d\omega}&=&- \frac{\alpha_sC_F}{\pi}{\rm Re}\int _{t_0}^{L+t_0} dt\frac{i}{\pi}(-2i\frac{\omega c_1(L+t_0,t)}{ D(t_1,t)}(1-e^{-\omega_1^2D(L+t_0,t)/(2\omega c1(tL+t_0,t))})\nonumber\\[10pt]
  &-&\theta_0^2\omega^2e^{\frac{\theta_0^2\omega^2D(L+t_0,t)}{2\omega c_1(L+t_0,t)}}
  (Ei(-i\frac{(\theta_0^2\omega^2+\omega_1^2)D(L+t_0,t)}{2\omega c_1(L+t_0,t)})-Ei(-i\frac{(\theta_0^2\omega^2)D(L+t_0,t)}{2\omega c_1(L+t_0,t)})))\nonumber\\[10pt]
  &\times&\frac{e^{-i\theta_0^2\omega (L+t_0-t)/2}}{\omega c_1(L+t_0,t)^2}.\nonumber\\[10pt]
  \label{11}
  \end{eqnarray}
  The function Ei is an Exponential Integral function \cite{Stegun}.
  The energy $\omega_1$ is the upper limit for the integration in $k_t$. In the energy loss calculations we have $\omega_1=\omega$ in ASW approach.
  On the other hand in the BDMPSZ  approach, as we mentioned above,  we just need to take the $\omega_1=\infty$ limit , and then subtract the vacuum contribution for regularization.   
\begin{eqnarray}
 \omega\frac{dI^{\rm bulk}}{d\omega }&=&-\frac{2\alpha_sC_F}{\pi}{\rm Re}\int^{L+t_0}_{t_0}dt\int ^{L+t_0}_tdt_1 (\frac{1}{D(t_1,t)^2}-\frac{1}{(t_1-t)^2})e^{-i\theta_0^2\omega(L+t_0-t)/2},\nonumber\\[10pt]
  \omega\frac{dI^{\rm boundary}}{d\omega }&=&- \frac{\alpha_sC_F}{\pi}{\rm Re}\int _{t_0}^{L+t_0} dt\frac{i}{\pi}(-2i\frac{ c_1(L+t_0,t)}{ D(L+t_0,t)}\nonumber\\[10pt]
  &-&\theta_0^2\omega e^{\frac{\theta_0^2\omega^2D(L+t_0,t)}{2\omega c_1(L+t_0,t)}}
  \Gamma(0,i\frac{(\theta_0^2\omega^2)D(L+t_0,t)}{2\omega c_1(L+t_0,t)}))
  \frac{e^{-i\theta_0^2\omega (L+t_0-t)/2}}{c_1(L+t_0,t)^2}\nonumber\\[10pt]
   &+&\frac{\alpha_sC_F}{\pi}{\rm Re}\int _{t_0}^{L+t_0} 
   dt\frac{i}{\pi}e^{-i\theta_0^2\omega (L+t_0-t)/2}(-2i/(L+t_0-t)\nonumber\\[10pt]
   &-&\theta_0^2\omega e^{i\theta_0^2\omega (L+t_0-t)/2}\Gamma(0,i\theta_0^2\omega (L+t_0-t)/2)),\nonumber\\[10pt]
    \label{12}
\end{eqnarray}
where $\Gamma (0,x)$ is an incomplete Gamma function \cite{Stegun}. Here the divergences in the integrals are regularized by the, subtraction
of the vacuum contribution that corresponds to the $\hat q_0\rightarrow 0$ limit.

\par It is easy to see that for sufficiently large $\omega$ the results of BDMPSZ and ASW coincide. Only for relatively 
  small emitted gluon energies  the results start to differ , see Fig. 1.
\begin{figurehere}
\centering
\includegraphics[scale=0.6]{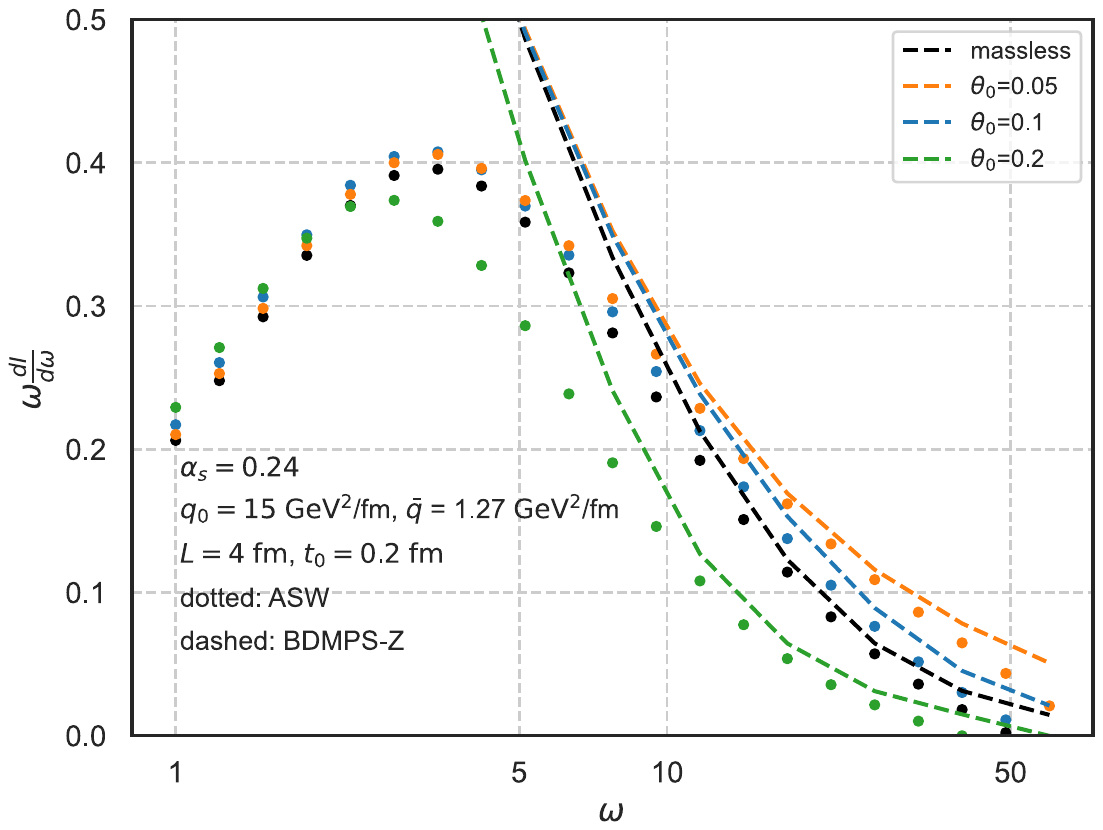}
\label{fig1a}
\caption{Energy distribution for expanding media in ASW and BDMPSZ approximations.}
\end{figurehere}

  \section{Salgado--Wiedemann scaling.}
  \par We shall now consider scaling. The scaling behaviour first found in \cite{SW1} means that the energy 
  loss is the function of only ratio $\omega/\omega_c$, i.e. the same as for the equivalent static configuration with
  \beq
  \hat q_{\rm stat}=\frac{2}{L^2}\int ^{L+t_0}_{t_0}(t-t_0)\hat{q}(t) dt.
  \eeq

  In particular for the linear expansion h=1 we have a simple expression:
  \beq
  \hat q_{\rm stat}=\frac{2\hat q_0}{L}t_0(1-\frac{t_0}{L}\log{\frac{L+t_{0}}{t}})\sim 2\hat q_0t_0/L=2\hat q(L),
  \eeq
  where the second equality is valid in $L>>t_0$ limit which is used in this paper.
  \begin{figure}[htbp]
  \begin{center}
\includegraphics[scale=0.38]{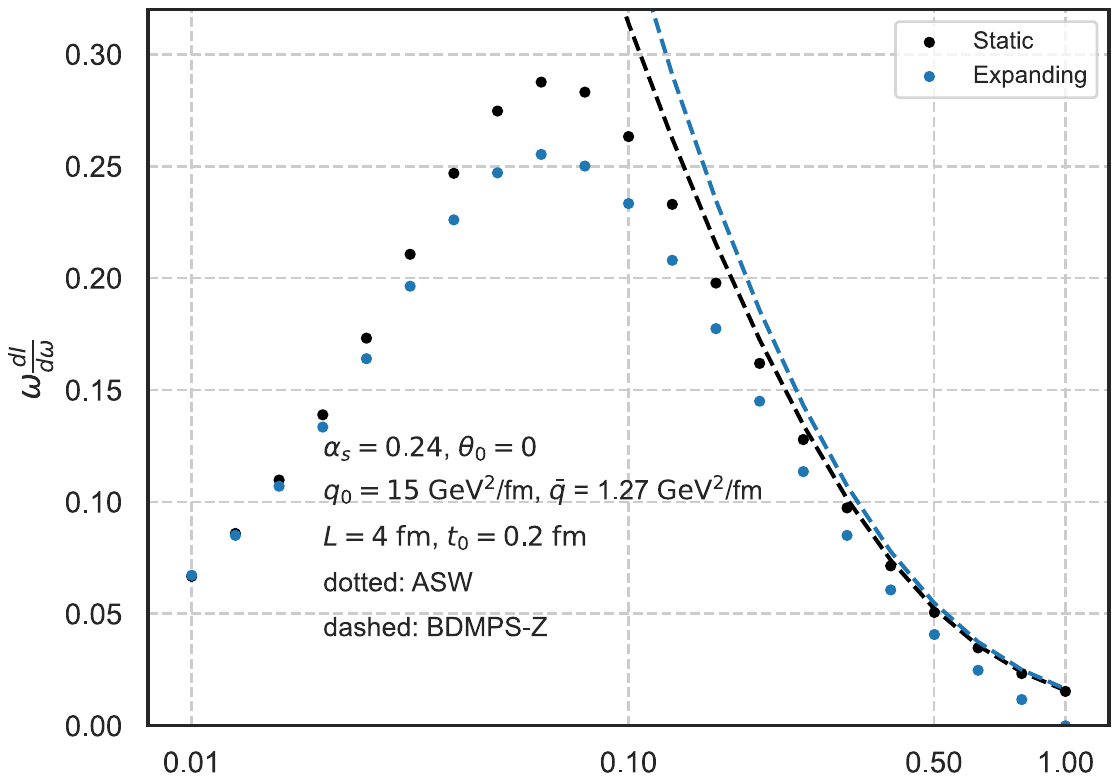}
.\includegraphics[scale=0.38]{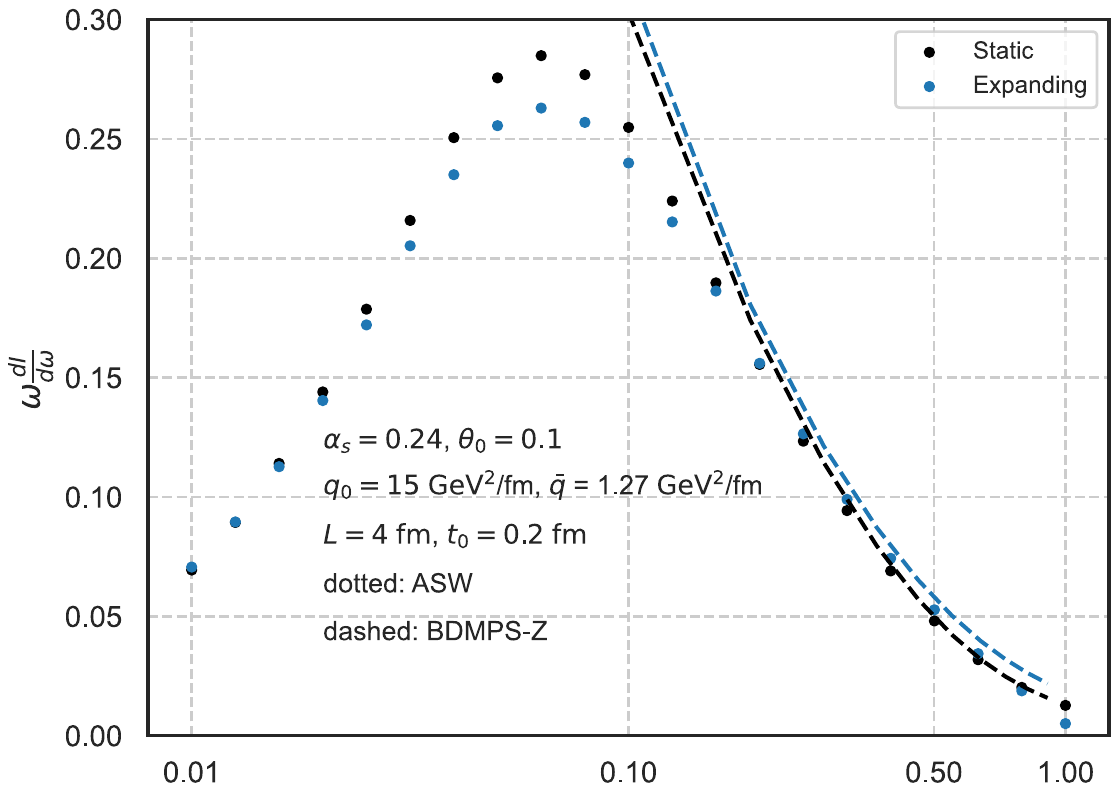}\\
\includegraphics[scale=0.38]{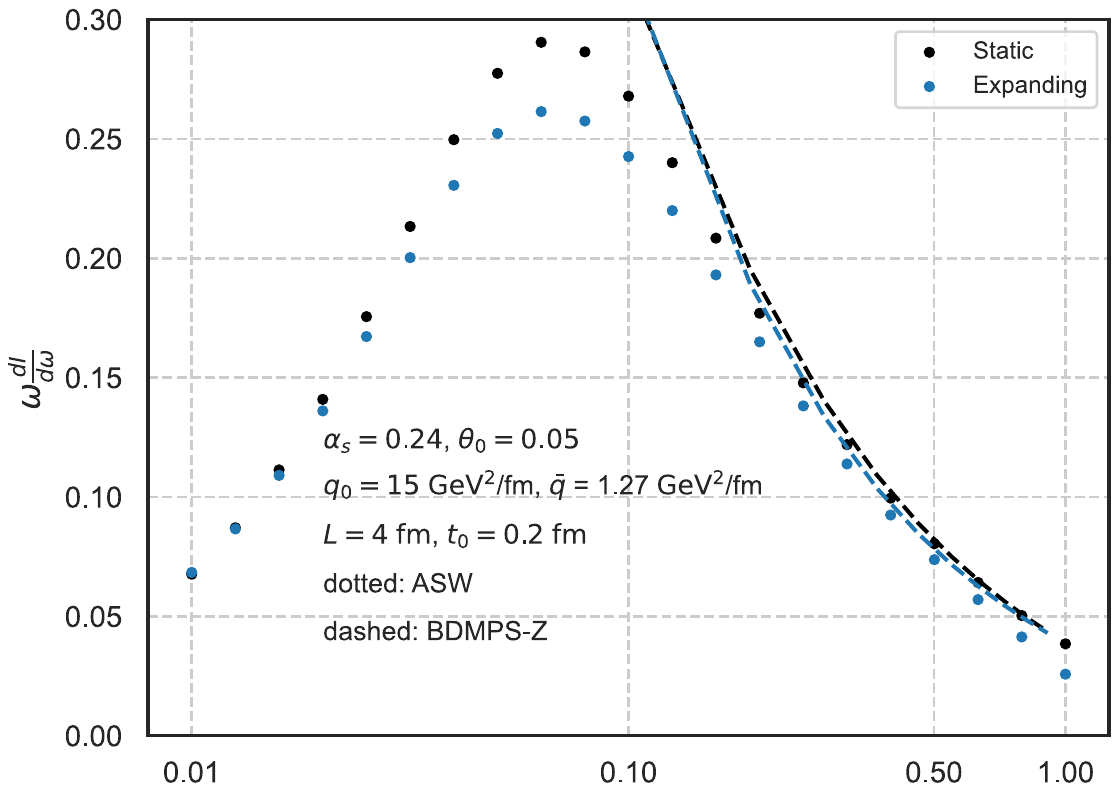}
\includegraphics[scale=0.38]{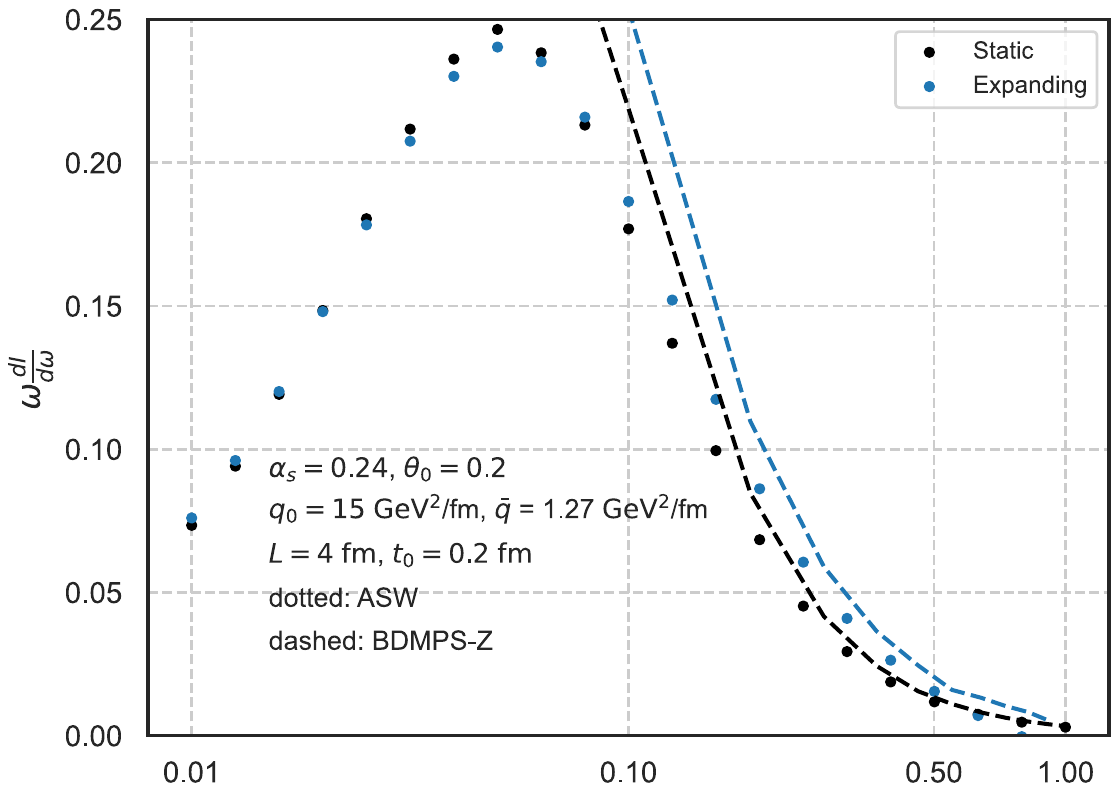}
\label{fig2a}
\caption{scaling: upper left-- the massles case,  down left $\theta_0=0.05$,upper right--$\theta_0=0.1$,down  right-$\theta_0=0.2$,}
\end{center}
\end{figure}
\par  let us denote
$ \omega_c=\hat q L^2/2$ for static media, and 
\beq
\omega_c=\int^{L}_{0} dt(t-t_0)\hat q_0(t_0/t)^h\sim \hat q_0Lt_0.
\eeq
for linearly expanding media.
\par We show in Fig. 2 the calculation results for the radiation spectrum as the function of $\omega$ for $t_0=0.2$ fm,$\hat q_0=3 $ GeV$^3$, however the same results are obtained 
for all other values of $\hat q_0$, and $0.1$ fm $\le t_0\le 0.2 $ fm. We see the good agreement between the energy radiation  spectrum 
for expanding and equivalent static , especially for $\omega\lesssim \omega_c$ \cite{SW2,T4,T3,T1,T2}. As it was mentioned in 
these references, and seen also in Fig. 2 the agreement is less good for small $\omega$, but these  small $\omega$ do not 
contribute to the energy loss of the parton (in distinction for the jet energy loss, as it is discussed in the section).
 \par The parton energy loss is usually characterised by the quenching weights. We refer to Ref. \cite{BDMPS5}
  for the detailed discussion of the quenching weights.
  We  assume the asymptotics for the tail of vacuum radiation:
  \beq
  \frac{d\sigma^{vac}(P_T+\epsilon)/d^2p_Tdy}{d\sigma^{vac}(P_T)/d^2p_Tdy}\sim \exp{(-n\epsilon /p_T)}.
  \label{31}
  \eeq
  The experimental measurements give $n\sim 5$ \cite{B3}, and $p_T $  is the energy of the leading parton/jet, and the $\epsilon$ is the radiated energy.
  \par Then the quenching weight is 
  \beq
  S(E)=\exp\left[-\int_{0}^{E}d\omega(1-e^{-n\omega/E})\frac{dI}{d\omega}\right].
  \eeq

\begin{table}[!h]
\begin{center}
\begin{tabular}{ |p{1.5cm}||p{1.cm}|p{1.cm}|p{1.cm}|p{1.cm}|p{1.cm} |p{1.cm}|p{1.cm}|p{1.cm}|p{1.cm}|p{1.cm}|p{1.cm} |p{1.cm}|}
 \hline
 & \multicolumn{6}{|c|}{Expanding}& \multicolumn{6}{|c|}{Static}\\
 \hline
 & \multicolumn{3}{|c|}{bottom quark}& \multicolumn{3}{|c|}{massless quark} & \multicolumn{3}{|c|}{bottom quark}& \multicolumn{3}{|c|}{massless quark} \\
 \hline
 $p_T$ GeV & 100 &  50 & 25  &100 &  50 & 25& 100 &  50 & 25  &100 &  50 & 25 \\
 \hline
ASW
 & 0.84   &0.79& 0.79 &  0.89.&0.8&  0.73
 & 0.83   &0.78& 0.8 &  0.87.&0.77&  0.7\\
 \hline
BDMPSZ
 & 0.8  &0.69& 0.64 &  0.78&0.65& 0.5
 & 0.79  &0.71& 0.69 &  0.79&0.62& 0.46  \\
\hline
\end{tabular}
\caption{\label{tab:multi}Quenching weights for b-quark, $m_b = 5$ GeV relative to massless quark}
\end{center}
\end{table}

\par For illustration we calculate in Table 1 the quenching weights  for expanding and equivalent static media
using the ASW and BDMPSZ formalisms.
 We see that  the quenching weights are very close.
 
   \section{Caucal-Iancu-Soyez  scaling.}
   
  \par As it was mentioned above, the scaling discussed in the previous section is relevant  for  the total energy loss of the parton, and works 
   for the emitted radiation frequencies $\omega\lesssim \omega_c$. Another type of scaling was suggested in \cite{T4,T3},
  see also Refs. \cite{T1,T2}, for the energy loss of massless gluon and quark jets. 
 \par This second type of the scaling is quite different from the scaling discussed in the  previous section that was valid in both ASW
 and BDMPSZ approaches. This  new type of scaling  is valid only for rather small values of $\omega$. It is based on the observation that the single parton spectrum  in the limit $\omega\rightarrow 0$ 
in the BDMPSZ approach  is  universal. It is  always proportional to $1/\sqrt{\omega}$:
  \beq 
  \omega\frac{ dI}{d\omega}=\frac{\alpha_sC_F}{\pi}\frac{2}{2-h}\sqrt{\frac{\hat q(L)L^2}{\omega}}.
  \label{a1}
 \eeq 
  This behaviour can be obtained analytically \cite{T3} for massless partons using the  explicit formula  for  massless gluon energy spectrum \cite{Arnold}:
     
    \begin{equation}
    \omega\frac{dI}{d\omega}=\frac{\alpha_s}{\pi}P(z)\log{\sqrt{\frac{t_0}{t_0+L}}\vert
    \frac {J_\nu(z_0)Y_{\nu-1}(z_L)-Y_{\nu}(z_0)J_{\nu -1}(z_L)}{_\nu(z_L)Y_{\nu-1}(z_L)-Y_{\nu}(zL0)J_{\nu -1}(z_L)}\vert},
    \label{as1}
    \end{equation} 
   and   the known asymptotics  of the Bessel functions J(x) and Y(x)  \cite{Stegun}.  Here 
   \begin{eqnarray}
    z(t)&=&\frac{2}{2-h}\frac{1+i}{2}\sqrt{\frac{\hat q}{\omega} }t_0(\frac{t}{t_0} )^{1-h/2}\nonumber\\[10pt]
   z_0&=&z(t_0),z_L=z(t_0+L).
   \label{ar2}
    \end{eqnarray}
    The equivalent static media quenching coefficient is then 
    \beq
    \hat q_{\rm stat}=(\frac{2}{2-h})^2\hat q_0(t_0/L)^h,
    \label{static}
    \eeq
    Here
    \beq
    P(z)=(1+(1-z)^2)/z
    \eeq
    is the corresponding DGLAP kernel for quark emitting the gluon, $\omega=zE$, where E is the jet energy.
  \par  It is easy to see that the singular part of the energy spectrum, i.e.  the term proportional to $1/\sqrt{\omega}$, 
   does not change if we include the mass of the radiating quark.  The reason is that the soft gluons described by scaling  are radiated at the angles much larger than the dead cone angle
  $\theta_0$.
The characteristic angle of radiation with frequency $\omega$ is $\theta_{\rm BDMPS}=(\hat q/\omega^3)^{1/4}$. The nonzero quark mass influences the radiation only when $\theta_{\rm BDMPS}\le \theta_0=m/E$ \cite{DK}, i.e.     
     $\omega>(\hat q E^4/m^4)^{1/3}\equiv (\hat q/\theta_0^4)^{1/3}$. For 
     typical static media  with $\hat q =0.3-0.6$ GeV$^3$, we get $\omega \ge 5-7$ GeV, that is rather far from soft gluon      
    region. For the expanding media the same arguments are valid, we just need to use the quenching coefficient at $t=L$--$\hat q(L)$ .
 \begin{figurehere}
\begin{center}
\includegraphics[scale=0.55]{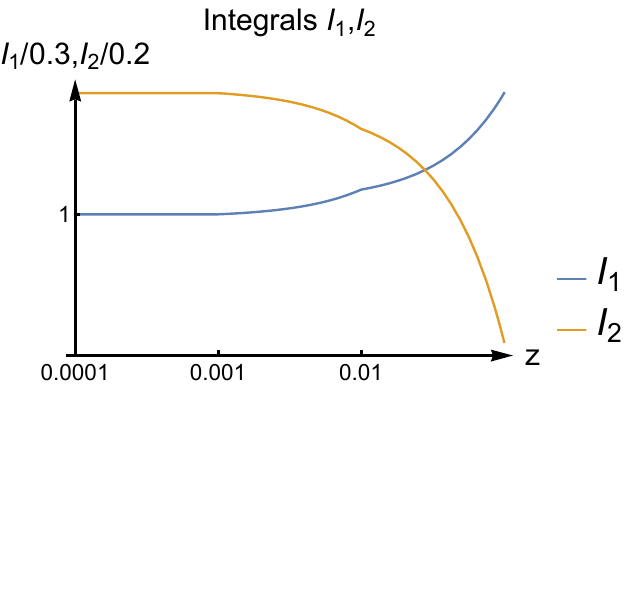}
\label{fig3b}
\caption{The function $I_1,I_2$ in the $z\rightarrow 0$  limit. Both functions tend to  constants in the $z\rightarrow 0$ limit.
we use h=1,$\hat q_0 =1$CeV$^3$.}
\end{center}
\end{figurehere} 
 
    \par From the mathematical point of view, consider Eq. 14. It is easy to see that the expansion in terms of the dead cone angle
    has the form
    \beq
   \omega \frac{dI}{d\omega}=\omega\frac{dI_0}{d\omega}+I_1\omega L\theta_0^2+I_2\omega^2L^2\theta_0^4`+..
   \label{mur}
   \eeq
   where the integrals $I_1,I_2,...$  are finite in the $\omega\rightarrow 0$ limit, and tend to a constant of order 1.
   This means that the asymptotics in the limit $\omega\rightarrow 0$ given by Eq. \ref{a1}  is not changed  by the quark mass.
  \par    Consider for example  the bulk contribution in \ref{mur}. This contribution is naturally expanded in the series 
    \beq \int^L_0dt_1\int^{t _1}_0 dt_1dt F(t_1,t)\exp(i\theta_0^2\omega(t_1-t) ) =\sum_{n=0}^{\infty} \theta_0^{2n}\omega^n\int^L_0dt_1\int^{t _1}_0 dt_1dt F(t_1,t)(t_1-t)^ni^n/n!,
  \eeq
  where n=0 corresponds to  the massless case.
    We are interested in the real part for even n and in the imaginary part of these integrals for  odd n. The corresponding  dependence on $z=\omega/E$
    in the limit $z\rightarrow 0$,
    for example of $I_1,I_2$, is depicted in Fig.3.  We see that indeed $I_1,I_2$ tend to constants.
    We have checked the higher coefficients $I_n$ in the $z\rightarrow 0$, and they are also nonsingular.
    Similar arguments can be made to show that the expansion of the boundary contribution to Eq. \ref{mur}  is also nonsingular.    It is also easy to check numerically that the difference between the massless and the massive spectrum goes smoothly to zero in the 
    limit $z\rightarrow 0$. 
    \par Thus we conclude that the singular part of the BDMPSZ spectrum is not influenced by the heavy quark mass,
    and the scaling \ref{static} is valid independently  of the dead cone angle value, with the quenching coefficient for for the equivalent static media given by Eq. 23.

    \par The scaling discussed above is,   strictly speaking  valid only in $\omega\rightarrow 0$ limit. It is violated for rather small $\omega$, although the agreement is rather good up to
    $\omega \sim \omega_{\rm br}$,
    as it was checked explicitly in \cite{T4,T1,T2}.  
    Here 
    \beq\omega_{\rm br}={\alpha_s^2c_Vc_F}{\pi^2}\omega_c
    \label{br1}
    \eeq
    and for massless partons this is physically the boundary of the region where the multiple radiation must be taken into account.
    \par The importance of this type of scaling is due to its connection to the jet energy loss. Indeed,  it was realised in
    \cite{turb}, that the energy loss of the jet  is dominated by soft gluons, that are radiated at large angles beyond the jet radius R..
       We can estimate the dependence of the jet energy loss on the dead cone value $\theta_0$ in a simple qualitative model.
       
    \par Recall  \cite{turb} that the energy loss due to media-induced radiation (MIE) of a jet of radius R  is given by a sum 
  \beq
  \Delta E =E_{\rm turb}+E_{\rm spec}.
  \label{4.1}
  \eeq
  Here  the first term corresponds to the multiple radiation of soft gluons at large 
  angles ( the turbulence cascade), larger than R:
  \begin{equation}
  E_{\rm turb}=
  E\left(1-e^{-\frac{v\omega_{{\rm br}}}{E}}\right)\sim v\omega_{\rm br},
  \label{turbo}
  \end{equation}
  where the second equality holds for $E>>\omega_{\rm  br}$, and 
  $v$ is a numerical constant  known from \cite{turb,Fister} that can be obtained by the numerical solution of the gluonic cascade equations.  
  \par The results of \cite{turb,Fister} mean that    the turbulent part of the jet energy loss  for  jets with $R<\theta_c/\alpha_s^2$,
  where $\theta_c$ is the coherence angle,
   will in a good approximation
  obey scaling laws \cite{turb,T4}. The energy loss of these jets can be calculated using eq. \ref{turbo}. 
  \par The second term in Eq. \ref{4.1}  the $E_{\rm spec}$is the single gluon spectrum that contributes to the jet energy loss.
  This term corresponds to the energy loss at frequencies $\omega \ge \omega_{\rm br}$, where no resummations of soft 
  gluons is needed in the leading approximation.
The corresponding frequencies are given by  $\omega\le \bar \omega=(\hat q(L)/(16\alpha_s^2R^4))^{1/3}$
(see the detailed discussion in \cite{T4}).
  The corresponding single gluon contribution is \cite{turb}
  \begin{equation}
  E_{\rm spec}=\int^{\bar \omega}_{\omega_{\rm br}}\omega\frac{dI}{d\omega} d\omega
 \end{equation}.
Typically $\omega_{br}<\bar \omega<\omega_{DC}$.  
  \par Since the scale $\bar \omega$ is typically far beyond $\omega_{\rm br} $ ,i.e. outside the scaling  region,   we must use  explicit single gluon spectrum given by Eq. 14.

  \par We can now develop a simple qualitative model to find the dependence of the jet energy loss on the dead cone angle $\theta_0$.
 \par  In order to  take it into account qualitatively the dependence of the jet  energy loss due to the turbulent cascade on $\theta_0$
  , recall that in the scaling region , that corresponds to small $\tau$ the turbulent energy  flow
  can be written as 
  \beq
  P(\tau)\equiv \frac{dE(\tau )}{d\tau}=v\tau.
  \label{tau}
  \eeq
  Here    the variable $\tau$ is an effective time 
  variable in the soft gluon turbulent cascade $\tau=\alpha_s\sqrt{q/E}t$,  where $t $ is the light-cone time. 
  The constant v depends only on $\omega\rightarrow 0$ limit. In the simplified BDMPS resummation model  $v$=$2\pi$.
  In order  to determine this constant more explicitly one can solve the gluon cascade equation numerically. Then v=4.98 for jet energy $E<\omega_c$ and   
 v=3.8 for $E>>\omega_c$ \cite{turb,Fister}.
  
 The integration boundary is just 
  $\tau_c=\sqrt{\omega_{\rm br}}$. The energy $E(\tau)$ is  the energy contained in the flow modes at the effective time $\tau$.
  Integrating Eq. \ref{tau} we obtain the qualitative estimate \cite{turb} for the turbulence cascade 
  \beq
  E=v\tau_c^2/2=v\omega_{\rm br}.
  \label{turb1}
  \eeq
  
  Then we can estimate the dependence  of the jet energy loss due to the turbulent part of the flow, on the 
  heavy quark mass,  using the physical definition of $\omega_{\rm br}$:
    \beq
    \int_{\omega_{\rm br}}d\omega \frac{dI}{d\omega}\sim 1
    \label{br2}
    \eeq
    i.e. this is the characteristic frequency such that for softer gluons one needs to take into account the multiple emissions.
    For massless partons equations \ref{br1} and \ref{br2} coincide.
    Using Eq. \ref{br2} we estimate the dependence of $\omega_{\rm br}$ on the dead cone angle, and then use this value of 
    $\omega_{\rm br}$  to estimate the jet energy loss for finite mass.
     
  This dependence will naturally be quite mild. Indeed, for the massless case we have   $\omega_{\rm br}=(\alpha_s/\pi)^2N_cC_F\omega_c$ , where $\omega_c=\hat q_{\rm stat}L^2/2$, and this scale is typically much smaller than 
  the dead cone scale $\omega_{\rm DC}=(\hat q_{\rm stat}/ \theta_0^4)^{1/3}$. The inclusion of $\theta_0$ changes 
  $\omega_{\rm br}$ only mildly, since the relevant soft gluons are radiated at angles larger than the dead cone value $\theta_0$. The corresponding values of $\omega_{\rm br}$ as the function of dead cone angle are depicted 
  in Table \ref{tab:wbr}.
 \begin{table}[!h]
\begin{center}
\begin{tabular}{ |p{1.5cm}||p{1.cm}|p{1.cm}|p{1.cm} |p{1.cm}|}
 \hline
 $\theta_0$ & 0 & 0.05 &  0.1 & 0.2 \\
 \hline
 $\omega_{\rm br}^{\text{\rm BDMPSZ}}$
 & 3.75 &3.75& 3.5 &  2.5\\
\hline
\end{tabular}
\caption{\label{tab:wbr}$\omega_{\rm br}$ for heavy flavour quark jet with $\alpha_s$ = 0.24, R = 0.4, L = 4 fm, $t_0=0.2$ fm,
$\hat q_0=3$ GeV$^3$, $\hat q_{\rm stat}$=0.6 GeV$^3$}
\end{center}
\end{table}
\par We assume that the jet energy is less than $\omega_c$, so we use v=4.98, while estimating the energy loss in Fig. 4.
\par 
  The results for the full energy loss of the jet as a function the dead cone angle are depicted   in Fig. 4.
  \par The determination of the dead cone dependence of the single gluon spectrum $E_{\rm spec}$ is straightforward--we just 
  use the explicit formulae for the energy spectrum of massive quark.
 \par The full quantitative solution of the problem of finding the $\theta_0$ dependence, including the corrections to scaling will demand the  
  the solution of the cascade equations. This will be however much more complicated problem than for massless 
  case, since for the massive case we do not know the analytical solution for the kernel of these equations, similar to
  the simple analytical solution given by Eq.  \ref{as1}.
  \begin{figurehere}
\begin{center}
\includegraphics[scale=0.65]{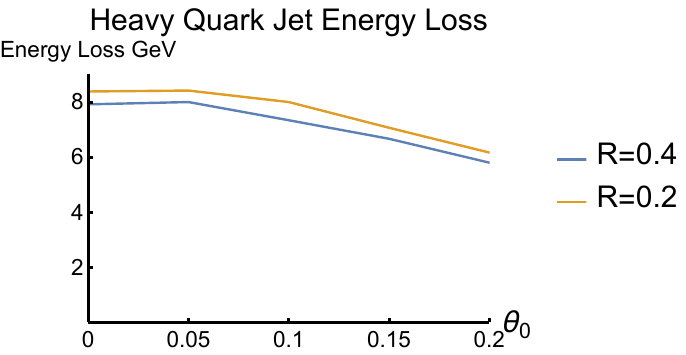}
\label{fig3a}
\caption{The average energy loss as a function of the jet energy/dead cone angle $\theta_0=m_b/E$   for heavy flavour jets  with radius R=0.2,0.4, $\hat q_0=3$ GeV$^3$,$t_0=1$GeV$^{-1}$.
}
\end{center}
\end{figurehere}         
        
     \section{Conclusions.}
 \par The analogous results are obtained for all other parameter ($\hat q_0$, $t_0,L)$values.
 We see that indeed scaling occurs for heavy quark  and heavy quark jet in the  expanding media with the quenching coefficient for effective static media $\hat q_{\rm stat}$ being independent 
 of dead cone angle,  both for the total heavy quark energy loss-Salgado-Wiedemann scaling
 \cite{SW1,SW2} and for the heavy quark jet energy loss \cite{T4,T3}.
 We have qualitatively estimated the influence of the dead come angle value on the jet energy loss.
 \par Finally let us note that the turbulent gluonic cascade and the corresponding energy loss were calculated in the BDMPSZ approach. If we take into account the phase constraints as in ASW, the spectrum goes down at the frequencies of the
 same order as $\omega_{\rm br`}$. That is $dI/d\omega$ is always of order 1-2 in the cascade due to quark radiation and
 naively there is  no need for resummation.
 This does not mean however that there is a contradiction in the soft region between ASW and BDMPSZ approaches..
 Indeed, the current approach to the gluonic cascade does not include back reaction, i.e. radiation by emitted gluons back inside 
  the jet cone. These higher order corrections  may lead to significant decrease of the jet energy  loss . (A somewhat related problem of backreaction to single gluon spectrum was considered in \cite{MT} for massless jets and extended to heavy quarks in \cite{B3}).  This subject deserves further studies.

 \par {\bf Acknowledgement} This research was supported by BSF grant 2020115.

 \end{document}